\documentclass{aastex}

\begin{document}

\title{The Oldest Stars of the Extremely Metal-Poor Local Group Dwarf Irregular
Galaxy Leo~A\footnotemark[1]}

\author{Regina E. Schulte-Ladbeck}
\affil{University of Pittsburgh, Pittsburgh, PA 15260, USA}
\email{rsl@phyast.pitt.edu}
\author{Ulrich Hopp}
\affil{Universit\"{a}tssternwarte M\"{u}nchen, M\"{u}nchen, FRG} 
\email{hopp@usm.uni-muenchen.de}
\author{Igor O. Drozdovsky}
\affil{University of Pittsburgh, Pittsburgh, PA 15260, USA, and University of
St.~Petersburg, St.~Petersburg, Russia}
\email{dio@phyast.pitt.edu}
\author{Laura Greggio}
\affil{Osservatorio Astronomico di Bologna, Bologna, Italy, and
Universit\"{a}tssternwarte M\"{u}nchen, M\"{u}nchen, FRG}
\email{greggio@usm.uni-muenchen.de}
\author{Mary M. Crone}
\affil{Skidmore College, Saratoga Springs, NY 12866, USA}
\email{mcrone@skidmore.edu}

\footnotetext[1]{Based on observations made with the NASA/ESA Hubble 
Space Telescope obtained from the 
Space Telescope Science Institute, which is operated by the Association of 
Universities for Research in Astronomy, Inc., under NASA contract NAS 5-26555.}

\begin{abstract}

We present deep Hubble Space Telescope single-star photometry of Leo~A 
in B, V, and I. Our new field of view is offset 
from the centrally located field observed by Tolstoy et al. (1998)
in order to expose the halo population of this galaxy. 
We report the detection of metal-poor red horizontal branch stars, which 
demonstrate that Leo~A is not a young galaxy. In fact, Leo~A is as least as 
old as metal-poor Galactic Globular Clusters which exhibit red horizontal branches,
and are considered to have a minimum age of about 9~Gyr. We discuss the distance to
Leo~A, and perform an extensive comparison of 
the data with stellar isochrones.  For a distance modulus of 24.5, the 
data are better than 50\% complete down to absolute magnitudes of 2 or more. 
We can easily identify stars with metallicities between 
0.0001 and 0.0004, and ages between about 5 and 10~Gyr, in their post-main-sequence phases,
but lack the detection of main-sequence turnoffs which would provide
unambiguous proof of ancient ($>$10~Gyr) stellar generations.   
Blue horizontal branch stars are above the detection limits,
but difficult to distinguish from young stars with similar colors and magnitudes. 
Synthetic color-magnitude diagrams
show it is possible to populate the blue horizontal branch in the halo of 
Leo~A. The models also suggest $\approx$50\% of the total astrated mass in our pointing to be
attributed to an ancient ($>$10~Gyr) stellar population. We conclude
that Leo~A started to form stars at least about 9~Gyr ago. Leo~A exhibits an
extremely low oxygen abundance, of only 3\% of Solar, in its ionized interstellar medium.
The existence of old stars in this very oxygen-deficient galaxy illustrates that a 
low oxygen abundance does not preclude a history of early star formation. 

\end{abstract}


\section{Introduction}

The ages of dwarf galaxies provide an important test of galaxy formation
models.   Galaxy formation through bottom-up gravitational collapse 
implies that small halos tend to form earliest. 
While many merge to form more massive halos,
small halos which have escaped merging may still be found today, and
in large numbers (e.g. Klypin et al. 1999, White \& Springel 2000,
Marzke \& Da Costa 1997, Ellis 1997). 
As the surviving building blocks of the large galaxies, today's
dwarf galaxies may have been hosts of the earliest star formation
in the Universe.

But there are also reasons one might expect to see nearby dwarf
galaxies with no old stars.  
First, low-mass halos in the early Universe were not necessarily effective 
at collecting gas, or at allowing gas to cool and experience
star formation. For instance, star formation in low-mass halos
may be effectively quenched by supernovae from the first burst, or delayed 
by the high UV background at re-ionization  (e.g. Dekel \& Silk 1986, Babul \& Ferguson 1996, 
Ferrara \& Tolstoy 2000, Barkana \& Loeb 2001).  Second, some 
small-scale density perturbations may have become non-linear only
recently, furnishing additional young, late-formed halos in the present epoch
(e.g. Roukema et al. 1997). Theory thus predicts a wide range 
of formation ages for the first stars in dwarf galaxies.

There is yet another possible formation history for dwarf galaxies. 
The more metal-rich, tidal dwarfs appear to result from galaxy mergers
(e.g. Duc \& Mirabel 1998), and can therefore form over an extended time period.  

On the observational side, it
is already clear that the dwarf galaxies of the Local Group have
experienced diverse star-formation histories (SFH) (e.g. Mateo 1998, 2000).
All well-studied dwarf Spheroidal (dSph) galaxies of the Local Group contain 
stars at least 10~Gyr old, as evidenced by main-sequence turn-off (MSTO) stars
with luminosities similar to those in old Galactic globular clusters (GC), the presence
of RR~Lyrae variables, and the presence of extended horizontal branches (HB) (e.g. Mateo
2000). DSphs with extended HBs often exhibit a population gradient: the higher-mass,
red HB (RHB) stars are more concentrated toward the center while the lower-mass,
blue HB (BHB) stars dominate at large distances (Harbeck et al. 2001). 
However, whether or not ancient stellar populations are also
ubiquitous in actively star-forming dwarfs,
such as dwarf Irregular (dIrr) or Blue Compact Dwarf (BCD) galaxies (of which there
are no examples within the Local Group), is not yet clear. Population
gradients in the young stellar component are clearly seen in dIrrs (e.g. Minniti \&
Zijlstra 1996, Grebel 1999) 
and BCDs (e.g. Schulte-Ladbeck et al. 1999, Tosi et al. 2001), but only for the most nearby dIrrs 
does current instrumentation permit the direct observation of old stars.

Evidence for ancient stars in dIrrs, while as yet scarce, does exist. 
RR~Lyrae stars have been detected 
in IC~1613 (Saha et al. 1992). WLM contains a globular cluster whose 
color-magnitude diagram (CMD) is similar to those of old, Galactic GCs 
(Hodge et al. 1999). The field populations of IC~1613 and
of WLM exhibit a BHB (Cole et al. 1999, Rejkuba et al. 2000), and
are considered to be an ancient stellar substratum. 

In contrast, Leo~A (= UGC~5364 = DDO~69) might represent a delayed-formation 
dwarf galaxy.
Tolstoy et al. (1998) obtained a CMD of Leo~A with HST/WFPC2, and found that  
the dominant stellar population in their centrally located field of view is less than 
about 2~Gyr old. They cautiously termed Leo~A a
``predominantly young galaxy" in the Local Group. 
Confirming the existence of a truly young galaxy in the present epoch 
would have a great impact on our understanding 
of galaxy formation, and motivates the research presented in this paper.

Interestingly, Leo~A is also the most metal-poor (i.e. oxygen-poor) galaxy of the 
Local Group. After the BCDs I~Zw~18
and SBS~0335-032, it is the third most oxygen-poor star-forming galaxy known today 
(a place it shares with UGCA~292). Its interstellar medium
(ISM) has an O abundance of only 3\% of Solar 
(12+log(O/H) = 7.30, Skillman, Kennicutt \& Hodge 1989, van Zee, Skillman \& Haynes 1999), 
almost as low as the famous I~Zw~18, which is at 2\% of Solar.
At the same time, Leo~A is about a factor of 15---30 closer, allowing its stellar content to
be resolved to deeper limits. 

Tolstoy et al. (1998) presented 
HST/WFPC2 observations down to about 25.5 to 26 in B, V \& I. They found 
that the CMDs of Leo~A 
are consistent with a formation time less than 2~Gyr ago. 
Specifically, their arguments are:

\begin{enumerate}

\item A horizontal branch, the signpost of an old stellar population in
GCs and dSphs, is not observed.

\item The narrow red-giant branch (RGB) is consistent with a young (1---2~Gyr) age and
little prior enrichment (if there were any older stars that 
enriched the gas), and so is the ratio of red clump (RC) to RGB stars.

\item The red giants do not extend spatially beyond the main-sequence (MS) stars, so 
there is no ``Baade's red sheet", the classical morphological indicator of a Pop~II.

\end{enumerate}

Since a small (10\%), underlying old ($\geq$~10~Gyr) population could not be ruled out, 
the authors cautiously call Leo~A a ``predominantly young galaxy" in the Local Group. 

The unique nature of Leo~A justified additional, and much deeper, HST observations.
We requested to obtain single-star photometry down to V$\approx$28 in 
the $\it{halo}$ of Leo~A, for the following reasons. 
First, Tolstoy et al. observed on the main body of Leo~A. Their
Figs.~5 \& 6 show the CMDs of the main body, using broad B, V \& I bands. 
The B,V data are near the photometric limits where one would expect to see
the RHB. Their B,V and V,I CMDs show that
any BHB is completely confused by the young stellar population. A detection of the
HB stars may however be possible with deeper imaging and at large galactocentric distance.
At larger distances from the center, there might be less of a young stellar population, 
resulting also in less crowding and blending of stellar images. 
Second, Tolstoy et al. argued that there is no spatial segregation between the
MS and the RGB stars, based on the fact that MS stars are also seen in
the outer regions of their data. On its digitized sky survey (DSS) image, 
the outer isophotes of Leo~A
extend well beyond the area covered by the Tolstoy et al. observations. 
In fact, Demers et al. (1984) find that the Holmberg diameter of Leo~A is 7'.
A pointing at large distance might reveal ``Baade's red sheet", a population of
old red giants, in addition to the old HB stars\footnotemark[2]. 
Therefore, in order to clarify unambiguously whether or not Leo~A is a young galaxy, we 
obtained new HST/WFPC2 data in a field further removed from the young populations.
Based on the new data, we make the case that stars with ages of at leat 9~Gyr,
and possibly greater,  are
present in Leo~A. Therefore, its age of formation is similar to other 
galaxies in the Local Group for which deep single-star photometry
is available.

\footnotetext[2]{Dolphin et al. (2002) recently showed stellar density maps
based on gound-based photometry which indicate the red-giant-star distribution is
larger than that of bright blue plume stars.}

\section{Observations}

Figure~1 shows the locations of the Tolstoy et al. pointing and our pointing.
The position of Leo~A in RA, Dec (2000) is 09$^h$59$^m$26$^s$, +30$^o$44'47" 
(Cotton, Condon \& Arbizzani 1999).
The center of the Tolstoy et al. pointing was at 09$^h$59$^m$27$^s$, +30$^o$44'22",
whereas our pointing was centered at 09$^h$59$^m$33$^s$, +30$^o$43'40", and had a roll
angle which positioned much of the WF chips into the outskirts of
Leo~A. 

\subsection {The Data}

Our observations were gathered as part of GO program 8575. We obtained
dithered exposures in three filters, with total exposure times of 22,200s in F439W, 
and 8,300s each in F555W and F814W.

The data reduction included the usual post-pipeline reductions and used the zero 
points from the May 1997 SYNPHOT tables. We performed single-star photometry 
with DAOPHOT~II (Stetson 1994) to obtain photometry
in instrumental magnitudes (in the HST Vegamag system). We also
transformed them to ground-based Johnson-Cousins B, V, I magnitudes with the
help of Holtzman et al. (1995). The chips which we are most interested in
are the ones farthest away from the center of Leo~A; we performed ADDSTAR 
tests on chips WF3 and WF4, and found that completeness is high well down 
to the RC. Very few Galactic foreground stars are expected due
to the high latitude of Leo~A and the small field of view of the WFPC2
(see Tolstoy et al.). No internal extinction is apparent in 
color-color diagrams of our data. We adopt the foreground
extinction of Schlegel, Finkbeiner \& Davis (1998, and see NED), namely 
A$_B$=0.089, A$_V$=0.068, and A$_I$=0.040 (slightly different from that
used by Tolstoy et al.). In Fig.~2, we show CMDs in [(F439W-F555W)$_0$, 
F555W$_0$], [(F555W-F814W)$_0$, F814W$_0$],
and [(F439W-F814W)$_0$, F814W$_0$] for each individual WFPC2 chip. 

Most of the CMDs we discuss below use photometry transformed into B, V, and I. 
Therefore, we give in Figs.~3 \& 4 DAOPHOT photometric errors and completeness fractions 
in B, V \& I, for both the WF3 and WF4 chips.

The Tolstoy et al. HST photometry reached estimated errors of 0.2~mag for B$\approx$25.5,
V$\approx$26.5, and I$\approx$25.5. Our data go about 2~mag. fainter for similar
errors.

\subsection {CMD Characteristics}

We detect a total of 4747 stars in B and V; 4136 in V and I;
and 3704 in B, V, and I.  For comparison, 
in the Tolstoy et al. data set there are 2636 stars in B and V, 
and 7295 stars in V and I. Although our data go to deeper limiting magnitudes
than the Tolstoy et al. data, our [V-I, I] set nevertheless contains 
a much smaller number of point sources. 
This can be understood in terms of a lower stellar density 
toward larger galactocentric radii, also seen in Fig.~1.

Figure 2 presents the CMDs for each chip in three different filter combinations.
The CMDs of all four chips exhibit blue and red plumes. The blue plume
consists of MS stars of a wide mass range, blue supergiant (BSG), and blue-loop (BL) stars. 
The BL stars are core-He burning descendants of massive and intermediate-mass stars,
and are also evident as bright stars between the two plumes. Subgiant branch (SGB) stars 
of lower-mass contribute to widen the bottom of the blue plume.
The red plume contains very few luminous red supergiant (RSG) and bright or thermally-pulsing
asymptotic giant branch (AGB) stars, but a well-developed
RGB/AGB branch. Note the well-populated RC in all four chips and in all CMDs. 
The RC area of the CMDs contains a mix of core-He burners 
with intermediate and low masses. At its bright and blue edge, there is a contribution
from intermediate-mass BL stars, whereas at the faint and blue edge, there is a spur
of low-mass RHB stars. The RC area is in part superimposed on the RGB and AGB. 
These two branches consist of
intermediate- and low-mass H-shell burning stars which evolve up the RGB, as
well as double-shell burning intermediate- and low-mass stars which ascend for 
a second time after core-He exhaustion. 

The two plumes merge toward faint magnitudes. 
This is attributed to the detection of subgiant branches of intermediate-mass stars 
after core-H exhaustion plus scatter due to increasing photometric errors and 
blending for fainter sources.

\subsection {Population Changes with Radius}

Figure~2 illustrates that the stellar content of Leo~A varies with position 
(compare Fig.~1 for the locations of the chips relative to the body of Leo~A).
We concentrate on comparing starcounts on the WF chips;
starcounts on the PC chip, with its different area and sensitivity, are not 
directly comparable to those on the WF chips.

Fig.~2a best shows the variation in the MS component as a function of
galactocentric distance. The blue plume on WF2 is seen to be very well populated at bright magnitudes.
MS stars are much less abundant, and the brightness of the bulk of MS stars 
becomes fainter, on WF3 and 4. Fig.~2c illustrates that another relatively 
young stellar component, the BL stars, are quite
prominent on WF2, and readily apparent on WF4, but rare on WF3.
The morphology of the RC also changes. This is best seen in Fig.~2c.
On WF2 and 4, the RC has a contribution of stars brighter than 24$^{th}$ magnitude,
which make it appear round. On WF3, these bright stars are missing,
and the RC morphology is that of a narrow triangle with the long side defined by 
stars with colors of (F439W-F814W)$_0$ $\approx$ 1. This blue part of the RC 
is clearly made up of HB stars, the lowest-mass and hence oldest stars which
can be distinguished on our CMDs. 

The stellar population changes with radius can also be illustrated with the help
of chip-by-chip luminosity functions (LF). The changes in the young stellar content with
radius are best seen in the V-band LF. In Fig.~5a, we show the F555W$_0$ LF
for all stars with (F439W-F555W)$_0$ $\leq$ 0. The top of the MS, defined as the magnitude at which the 
F555W$_0$ LF first systematically rises above zero, occurs at 23.0$\pm$0.1 on WF2, 25.3$\pm$0.2 on WF4,
and 25.5$\pm$0.1 on WF3. The increase in stellar density just below magnitude 25
on WF3 \& WF4 could also be due to a large relative contribution by BHB stars.

Starcounts on the [(F439W-F555W)$_0$, F555W$_0$] CMDs along the top of the blue plume with limits
20 $<$ F555W$_0$ $<$ 25 (where the faint limit is set by the onset of a potential 
BHB contribution) and (F439W-F555W)$_0$ $\leq$ 0,
yield 194$\pm$14 bright blue stars on WF2, 24$\pm$5 on WF4, and 18$\pm$4 on WF3 
(the ``errors" are $\sqrt{N}$). This corresponds to relative numbers of bright, blue-plume
stars on the WF chips in proportions 1 : 0.12 : 0.09.
From Fig.~2b, we see that BL stars occupy the 
region 20 $<$ F814W$_0$ $<$ 24 and 0 $<$ (F555W-F814W)$_0$ $<$ 0.6. We find 48$\pm$7 stars
on WF2, 18$\pm$4 on WF4, and 7$\pm$3 on WF3, or BL stars in ratios of 1 : 0.38 : 0.15. 
The LFs and the CMDs thus indicate a strong gradient in the number density of young stars with distance
from the center of Leo~A. 

The changes in the intermediate-age/old stellar content with radius are best seen in the I-band LF. 
Fig.~5b shows chip-by-chip LFs for the red plume. They were constructed from the
[(F555W-F814W)$_0$, F814W$_0$] CMDs with color selection criterion 0.65 $<$
(F555W-F814W)$_0$ $<$ 1. A small, gradual decrease in stellar numbers from WF2, to WF4, to WF3, 
is apparent. This indicates changes in the numbers of RC and RGB/AGB stars, but at a much
smaller rate than the change seen in the young, MS component. Counting stars in the
top mangitudes of the RGB/AGB on the [(F555W-F814W)$_0$, F814W$_0$] CMDs 
with 19.5 $<$ F814W$_0$ $<$ 23 and 0.8 $<$ (F555W-F814W)$_0$ $<$ 2,
we find 54$\pm$7 on WF2, 53$\pm$7 on WF4, and 27$\pm$5 on WF3, or relative numbers of 1 : 0.98 : 0.5. 
When we define the area of the RC to encompass stars with 23 $<$ F814W$_0$ $<$ 24.5 and 
0.6 $<$ (F555W-F814W)$_0$ $<$ 1.2, we count 287$\pm$17 stars on WF2, 236$\pm$15 on WF4, and 185$\pm$14 WF3,
or ratios of 1 : 0.82 : 0.64. This supports the visual impression of a less rapid change 
with distance in the intermediate-age/old as compared to the young stellar component.

The RC also changes in morphology. This can best be seen in Figs.~2c. 
At the same time (as seen e.g. on Fig.~5b), the RC LF appears to be 
wider on WF2 than it is on WF4 and WF3. This suggests the RC on WF2 
has a larger contribution by comparatively young RC stars than the RC on WF3 and WF4. 
We selected stars on the [(F439W-F555W)$_0$, F555W$_0$] CMDs with 1.2 $<$ (F439W-F555W)$_0$  $<$ 1.6 and
23 $<$ F814W$_0$ $<$ 24.5. The starcounts in the RC are characterized by the
following mean F814W$_0$, sigma, and kurtosis: on WF2  23.89, 0.30, 0.09;
on WF4  23.93, 0.25, 0.42; and on WF3  23.98, 0.24, -0.23. 

Fig.~6 is a V, I CMD of all stars on all chips, which allows
for a comparison with the V, I CMDs of Tolstoy et al. (e.g., their
Figs.~6 and 16). Changes in the young stellar component with radius are
readily apparent. In the Tolstoy et al. CMD of the center field
of Leo~A, the top of the MS is seen at an I magnitude of about
23. This compares to an I magnitude of about 24 in our
data. The MS on the combined CMD is dominated by that of WF2,
on WF3 and WF4 the top of the blue plume is clearly fainter yet.
Star formation stopped at an earlier time in our outlying field,
compared to the center, where it was active more recently.
All of the Tolstoy et al. [V-I, I] WF chip CMDs show a
prominent spur of BL stars emanating from the top of the RC. In our data, the
BL stars are rather weak. They are prominent only on the WF2, weaker
on the WF4, and very rare on the WF3 CMD. Interestingly both
our V, I CMD and the CMD of Tolstoy et al. contain very few luminous red stars
which could be interpreted as RSG or bright AGB stars. 
Tolstoy et al. barely detected the RC in their [B-V, V] CMD.
In their V, I CMD, on the other hand, the RC is a strong, vertically extended feature, 
elongated along the magnitude direction. In our V, I CMD, the
RC has a component that extends horizontally, along the color axis.
This is the signature of a much older RC (see Caputo et al. 1995).
We interpret this to indicate that old RHB stars become more obvious 
with increasing distance from the center of Leo~A because 
the contribution of intermediate-age RC stars decreases 
as the mean population age increases.

Tolstoy et al. could not clearly distinguish HB stars in their V, I, or B, V CMDs, 
owing to the shallow limiting magnitudes which they were able to achieve. 
While we can distinguish RHB stars on our deeper CMDs, 
discriminating BHB stars against the MS component, which remains
strong even at the position of our pointing, is not possible. We place limits
on a possible BHB component later in this paper.

In summary, the positional variations in the MS and BL stars as well as the
changes in RC morphology and the detection of RHB stars, imply that the 
mean age of the stellar content of Leo~A becomes greater with increasing 
distance from the center.

\section{Results and Discussion}

In general, the quantitative analysis of CMDs in terms of SFH is conducted by comparing
the data with theoretical isochrones, tracks, or synthetic CMDs based on models
of stellar evolution and atmospheres. This requires knowledge of
the distance. Distance turned out to be an important parameter when Tolstoy et al.
attempted to deduce the SFH at the center of Leo~A. This problem remains
acute in the interpretation of our data, as we attempt to find a solution
which accommodates the magnitude of the TRGB, the color of the RGB 
and the location of the RHB stars. 

\subsection {Distance}

In this section, we derive the distance to Leo~A using the I-band TRGB.
We use this distance to conduct the comparison with theoretical models 
which follows in the next sections. We also discuss the distance on pre- and
post-Hipparcos distance scales. We emphasize that
while the specific ages in the next section depend on our adopted distance
scale, our main conclusions do not. 
 
Table~3 of Tolstoy et al. (1998) is a summary of distances
to Leo~A derived with different methods and assumptions. There exists
a wide range of distance moduli, from 23.9($\pm$0.1) to 24.9($\pm$0.7).
In principle, our observations of Leo~A 
offer two means for deriving its distance, namely the TRGB
and the RC. 

A well-defined TRGB is present in $>$1~Gyr old stellar populations of a wide range of 
metallicities, and the absolute I-magnitude of the TRGB can yield a distance
with a precision and accuracy similar to that of the Cepheid
method (Lee, Freedman, \& Madore 1993). A small metallicity correction may be
applied; and a metallicity determination can be made from the V-I color of the RGB
at either 0.5 or 1.0~mag below the TRGB (see Lee et al.
1993, also Bellazini, Ferraro \& Pancino 2001). As long as the metallicity 
is lower than [Fe/H]$\approx$-0.7, the TRGB in
the I band is observed to be a constant to better than 0.1~mag.
The location of the TRGB is commonly found by detecting a sharp
edge in the I-band luminosity function of the red plume based on a
[(V-I), I] CMD. In Fig.~6, we show the [(V-I)$_0$, I$_0$] CMD of our WFPC2
data. It is immediately obvious that the small
number of stars detected in our field of view prohibits a reliable 
determination of the TRGB magnitude, or of the V-I color near the TRGB. 
We show in Fig.~6 that our CMD and our I-band LF are consistent with
I$_{TRGB,0}$ = 20.5, which is the value that Tolstoy et al. derived using 
ground-based, wide-field imaging. Tolstoy et al. did not trust the TRGB as a 
distance indicator because their data are consistent with
a predominantly young stellar population. Our data, on the other hand, demonstrate that
stars with ages of at least 9~Gyr are present in Leo~A. We can ascertain that the data meet
both the requirements, of a low metallicity $\it{and}$ of a large age, which need to be satisfied
before using the TRGB as a standard candle. The field of view of the ground-based images
on which Tolstoy et al. based the TRGB is 9' x 9', overlapping both their,
as well as our pointing. This justifies our adoption 
of the Tolstoy et al. ground-based TRGB. 

In Fig.~7, we overplot the GC ridgelines of Da Costa \& Armandroff (1990) onto
our best V, I data, assuming I$_{TRGB,0}$ = 20.5
and hence, (m-M)$_0$$\approx$24.5, for Leo~A. 
As can be seen from Fig.~7, the metallicity of the RGB stars of Leo~A is
near that of M~15, which has [Fe/H]=-2.17. There are stars in Leo~A which
lie to the blue of M~15 GC ridgeline. Tolstoy et al. argued in part from this
fact that the distance modulus
of Leo~A must be shorter than that indicated by the TRGB. However, there is a range of other
explanations for these stars. They could be RGB stars with a metallicity
below that of M~15. Some could be from the old AGB.
And, since Leo~A has formed stars over an extended period of time, we cannot
exclude a young AGB or RGB, either. Tolstoy et al. do not
provide an error for their ground-based TRGB value, but judging from the data,
it appears that the error could be 0.1 -- 0.2~mag. Adopting a dispersion
for the metal-poor GC TRGBs of about $\pm$0.05 (ignoring NGC~6397), the TRGB
distance modulus of 24.5 for Leo~A has an error of about $\pm$~0.1 -- 0.2~mag.

The GC ridgelines of Da Costa \& Armandroff are on the ``old" RR~Lyrae
distance scale which results from theoretical HB models of Lee, Demarque \& Zinn (1990).
This distance scale changed in the post-Hipparcos era, but it is not
entirely clear what the new zeropoint is and how it depends on metallicity
(e.g. Popowski \& Gould 1999, Carretta et al. 2000).
If we assume the
calibration of Udalski (2000a), which corrects the TRGB to the RR~Lyrae calibration
of Popowski \& Gould (1999), then M$_I$=-3.91($\pm$0.05) for [Fe/H]$<$-0.7.
The distance modulus of Leo~A on this ``new" RR~Lyrae distance scale is
(m-M)$_0$=24.4. The error is the same as that derived above, about $\pm$~0.1 -- 0.2~mag. 
We note that adopting this distance modulus for Leo~A does
not change the conclusion that its RGB stars are extremely metal poor.

The average RC magnitude has been calibrated empirically
as a distance indicator. A discussion of its application to nearby galaxies
may be found in Udalski (2000~a,b). RC stars occupy some of the same space
as RGB/AGB stars on CMDs.
In order to find I$_{RC,0}$, these components are usually separated by
fitting the I-band luminosity function with two functions. We use a linear
one for the RGB/AGB contribution, and a Gaussian one for the RC contribution.
In Fig.~8, we show the red plume LF based on our best V, I data, with such
a fit overlayed. The color selection criterion is 
0.5$\leq$(V-I)$_0$$\leq$1.0. From our fit we find the position of the RC, 
I$_{RC,0}$ = 24.05$\pm$0.05, and its 1$\sigma$ width, 0.20$\pm$0.05.
If we assume M$_I$=(0.14$\pm$0.04) x ([Fe/H]+0.5) - 0.29$\pm$0.05 from
Udalski (2000a) and use [Fe/H]=-2.17 for the RC stars of Leo~A, 
then M$_I$ is -0.52 and the distance modulus for Leo~A is (m-M)$_0$$\approx$24.6.
Since we have neither determined a good value for [Fe/H] from our data,
nor derived an [Fe/H] error, we can only give a lower
limit to the distance modulus error from the RC method, $>$0.1~mag.
The distance modulus from the RC method is consistent with
that derived via the TRGB method.

Udalski worked on the assumption that the absolute I magnitude of the
RC has a small dependence on metallicity, and virtually none on age,
for 2---10~Gyr. This contrasts with theoretical work that suggests 
the absolute magnitude of the clump is a  
function of both age and metallicity
(Girardi \& Salaris 2001). Galaxies with ongoing
star formation in particular, are found to exhibit an age distribution
of clump stars which is strongly biased toward younger (1---3~Gyr) ages, and
the RC magnitude gives a false distance if this is ignored. 

Tolstoy et al. measured the RC at a mean I$_0$ magnitude of 23.73 (corrected for
A$_I$=0.04). This is 0.32~mag brighter than the RC we find in our data.
Tolstoy et al. indicate that in their determination they favored the peak of the I histogram and its
low-luminosity side over the high-luminosity side which they thought was obviously
contaminated by BL stars. We also favored the peak in Fig.~8. 
Measurement errors or incompleteness, and calibration uncertainties, 
could easily account for about 0.1~mag,
but are unlikely to be as large as several tenths of a magnitude. Assuming therefore
that this difference is real, an alternative explanation could be 
age (and metallicity) biasing following Girardi \& Salaris (2001). For example, if there is a lot
of power in the young RC, it can easily bias the RC of a composite stellar population
toward a bright absolute I magnitude. As further support of this
interpretation, we note that the weaker, secondary peak seen in Fig.~8 can be described by a
Gaussian with a peak I$_0$ magnitude of 23.65$\pm$0.05 and a width of 0.15$\pm$0.05.
This agrees with the peak RC magnitude found by Tolstoy et al.(1998). The differences
in RC magnitude between the Tolstoy et al. data and our data could therefore
be accounted for by the difference in pointing: the field of Tolstoy et al.
was more centrally located and contains a population with a younger mean age than our 
outlying field, biasing their RC to a brighter magnitude. 

As an aside we note that the age and metallicity 
dependence of the RC is still hotly debated between observers and theorists (e.g. Udalski 2000b). 
Leo~A, with its extremely metal-poor stellar content and core-halo population gradient, 
may serve as a good future test case because theory predicts that
the slope of M$_{RC,I}$ steepens 
with decreasing metallicity (Girardi \& Salaris).

In summary, the application of both the TRGB and RC methods to our WFPC2 data 
consistently indicates that the distance to Leo~A is m-M=24.5($\pm$$\approx$0.2)\footnotemark[3].
This value is larger than, but not inconsistent with, that of Tolstoy et al. (1998),
who chose to adopt m-M=24.2($\pm$0.2).

\footnotetext[3]{Dolphin et al. (2002) derived a distance to Leo~A based on their discovery of
RR~Lyrae variables with a mean V magnitude of 25.10$\pm$0.09. Using the calibration of
Carretta et al. (2000), and adopting [Fe/H] = -1.7$\pm$0.3, they determine a true distance modulus
of 24.51$\pm$0.12 for Leo~A.}

\subsection {Comparison with Isochrones}

Several databases for isochrones exist, but not all cover the metallicities and stellar phases 
which we need to interpret the Leo~A data. The stellar-evolution models are based on different
input physics, and different stellar atmospheres are adopted
to convert from the theoretical to the observational plane. 

We match the isochrones to the data by first
determining the average I magnitude at the TRGBs of the oldest
isochrones ($\geq$10~Gyr) in each database. We then shift the isochrones by
the difference between the observed apparent TRGB magnitude 
and the theoretical absolute 
TRGB magnitudes in the I band. We adopt the following absolute
I magnitudes at the TRGBs for the isochrones: for ``old" Padua (i.e., Bertelli et al. 1994) 
Z=0.0004 (or [Fe/H]=-1.74), M$_I$=-4.1, m-M=24.6;
for ``new" Padua (Girardi et al. 2000, and http://pleiadi.pd.astro.it/) Z=0.0004, 
M$_I$=-4.0, m-M=24.5; Z=0.0001 (or [Fe/H]=-2.36), M$_I$=-3.9, m-M=24.4;
for Yale (Yi et al. 2001) Z=0.0004, M$_I$=-4.1, m-M=24.6; Z=0.0001, M$_I$=-4.0, m-M=24.5; and
for Frascati (Cassisi, Castellani \& Castellani 1997) Z=0.0001, M$_I$=-4.1, m-M=24.6.
We note that all of the isochrones have I-band TRGBs which yield distances that
are consistent with our empirical distance estimate. We also note that the isochrones 
predict the location of the RC at different magnitudes; this is illustrated to some extent by the
figures in this section (also see Castellani et al. 2000, for a detailed discussion).

Tolstoy et al. (1998) modeled their Leo~A data with the old Padua models. 
The Z=0.0004 isochrones provided a good description of the young stellar content.
This is not surprising, 
because their metallicity is similar to the oxygen abundance 
of the ionized ISM of Leo~A. Since the work of Tolstoy et al., a new set of
Padua isochrones was published. We therefore provide in Fig.~9 a comparison of our data with both
the old and the new Padua isochrones. As can be seen from Fig.~9, the difference between
the old and the new Padua isochrones is rather small. 

We show three sets of isochrones which represent the young, intermediate-age, and old 
stellar content of Leo~A.
The densely populated area of the blue plume can be interpreted to contain young
MS stars. The MSTO of a 200~Myr isochrone matches well the top of the MS,
while the detection limits in the blue plume correspond to the 
MSTO of about the 2~Gyr isochrone. The less densely populated part of the blue plume 
above I$_0$$\approx$24 can be accounted
for with BSG and BL stars with ages of about 200~Myr and older. The few bright stars located between
the blue and red plumes can similarly be ascribed to BL stars with ages of a few hundred Myr.
Stars just blueward of the top of the RC are consistent with stars less than about 500~Myr
old which are in the core-He
burning phase. While the isochrones extend well above the TRGB, they predict a very small number
of luminous red stars, as observed.  
The bottom of the densely populated area of the blue plume might contain 
BHB stars with ages well in excess of 10~Gyr. The 18~Gyr isochrone suggestively connects the 
RHB stars to a handful of stars at I$_0$$\approx$25 just redward of the blue plume.
Alternatively, these stars could be just beyond the MSTO of a young, $\approx$1~Gyr
population. The 18~Gyr isochrone extends only slightly above the 
TRGB to include AGB stars.  
Indeed, few stars are observed here.

The RC area is characteristic of an intermediate-age population. The 2~Gyr
isochrone is shown to illustrate the upper age limit derived by Tolstoy et al. from their
centrally-located field. Indeed the bright,
red portion of the RC can be associated with ages of around 2~Gyr in our data as well. 
But a substantial number of stars with ages above 2~Gyr is
required to match the entire extent of the RC, and in particular, its
faint, blue portion. The 2~Gyr isochrone remains blueward of the RGB in both CMDs
(and so would a Z=0.0001, 2~Gyr Padua isochrone). 

Therefore, in order to explain the RGB stars of Leo~A with isochrones of metallicity Z=0.0004,
old ages are required, but even they cannot provide a perfect match. 
The 18~Gyr isochrone shown in
Fig.~9 stays slightly blueward of the TRGB in the [(V-I)$_0$, I$_0$] CMD, and is slightly red for
portions of the [(B-I)$_0$, I$_0$] RGB of Leo~A. The fact that the RGB can be explained with
ancient stars of such a high metallicity contradicts our results from using GC ridgelines. 
The empirical GC ridgelines indicate both old $\it{and}$ very metal-poor stars. Indeed,
it is well known that, being very sensitive to the stellar atmospheres adopted, 
the slopes of theoretical giant branches are difficult to predict in the observational plane.  

In Fig.~10, we overplot onto our [(B-I)$_0$, I$_0$] CMD
Yale isochrones of intermediate and old ages for Z=0.0004, and Z=0.0001. 
For Z=0.0004, a predominantely intermediate-age
RGB provides the best match to the data. For Z=0.0001 (slightly lower than the metallicity
of the M~15 RGB), on the other hand, even the oldest isochrones
are to the blue of the observed RGB. 
These isochrones suggest that for metallicities $Z\leq 0.0004$, there are
RGB stars with a minimum age of 5~Gyr. 

The Padua and the Frascati databases include the HB phase. In Fig.~11, we overplot 
the Leo~A [(B-I)$_0$, I$_0$] CMD
with intermediate-age and old isochrones from these databases.
We show Padua isochrones for Z=0.0004 which we know can fit the RGB/AGB; but
this time our goal is to investigate the RC/RHB. We find that they
are slightly too faint at the RC/RHB.  We also show Padua and Frascati isochrones for Z=0.0001. 
The Padua Z=0.0001 isochrones are too blue on the RGB; however, because they are bright
in the core-He burning phase, they can give a good characterization of the RC/RHB.
The 5~Gyr isochrone, for instance, is a good match to the RC stars. The 10~Gyr
isochrone extends slightly blueward of the data, indicating that 
the blue edge of the RHB is just under 10~Gyr of age. 
The Frascati Z=0.0001 isochrones match very well the RGB/AGB stars of Leo~A, and
suggest a population which is predominantly several Gyr old.
At the same time, the bluest RC/RHB stars can be accounted for with the 5 and 10~Gyr isochrones,
while the 14~Gyr isochrone extends beyond the blue edge of the RHB in the data. 
Since the location of HB stars also depends on how much mass loss is assumed on the RGB, we cannot
give a firm age limit based on the blueward extension of the RHB.

We summarize our findings as follows. Stars with a wide range of ages, between about 0.2
and 2~Gyr, make up the young stellar content of Leo~A within our field of view. 
Stars of such young ages cannot account for the location of the RGB and the RC in the data. 
A minimum age of about 5~Gyr is required for metallicities
between Z=0.0004 and 0.0001 in order to explain the color of the RGB/AGB in Leo~A. For similar
metallicites, the bulk of the RC stars is several Gyr old. 
The blueward extent of the RHB is consistent with ages of up to about 10~Gyr.

\subsection {Comparison with Synthetic CMDs}

The RHB overlaps partially with the RC and RGB/AGB, while the BHB coincides largely with 
MS stars. In order to investigate how much of a horizontal branch component 
is consistent with our CMDs of the outer regions of Leo~A, we here present synthetic 
CMDs of the stellar content in the WF3 \& WF4 chips. 

Synthetic CMDs, which are based
on theoretical stellar evolutionary tracks and atmospheres convolved with the photometric
errors and completeness fractions of the data, can provide a more complete picture of the
SFH than is possible with the isochrone comparison conducted in the previous section.
Here, we use the Bologna Code (Greggio et al. 1998) with the Padua tracks of 
metallicity Z=0.0004 (Fagotto et al. 1994), the same ones adopted by Tolstoy et al.
(1998), and the stellar atmospheres of Bessel, 
Castelli \& Plez (1998). The data errors and the recovery fractions from
false star tests on WF3 \& WF4 are shown in Figs.~3 \& 4. We incorporated photometric 
uncertainties and completeness fractions based on these results into the simulator. 
For reference, the distance modulus implied by the tracks is 24.6; at this distance,
the area of the two WF chips equals 0.19~kpc$^2$, and absolute model SFRs may be transformed
into SFRs/area accordingly.

It should be noted that the completeness tests were carried out in coarse bins; and this provides
a limitation to the comparison of the models and the data in the faintest magnitude bin near
the completeness limits. Furthermore, as shown by Fig.~3, there are a few outlying 
V and I errors compared with the broad band which defines most of the errors. The simulator
was programmed to contain the broad band of errors; but we did not attempt to
include the outliers. Therefore, we anticipate mismatches of the model with the data
for the faintest magnitude bin, and we expect the synthetic CMDs to lack a smattering
of objects with extreme colors. There are additional sources of error which are not
included in the simulations. Internal, differential reddening does not appear to
be important in Leo~A, and was neglected. Contamination by foreground stars also is
a very small source of error which we did not address with the simulations.

The simulations presented here assume a Salpeter (1955) initial mass function 
(IMF) with the standard slope of 2.35. Since many investigations of the SFRs and SFHs of
dIrr galaxies have assumed mass limits from 0.1 to 100~M$_\odot$, we do so as well.
The primary constraint of the modeling is that any linear combination of synthetic
star forming events has to produce 1738 surviving stars on the CMD with colors and 
luminosities which are in broad agreement with
the data. We followed the same methodoloy as Schulte-Ladbeck et al. (2000, 2001) and Crone et al. (2002),
in that we started by identifying boxes in color and magnitude which contain stars with
progressively older ages. We then modeled the stars in each box with a constant SFR.
For example, an appropriate constraint of the recent SFH within
the past few hundred Myrs can be obtained by considering the BL descendants found at colors of
about -0.4 $<$ (V-I)$_0$ $<$ 0.4 and 22 $<$ I$_0$ $<$ 23.5. The number of stars observed 
between about 0 $<$ (V-I)$_0$ $<$ 0.4 and 24 $<$ I$_0$ $<$ 25.5 can be used to constrain
the BHB stars, for which the isochrones have indicated extremely large ages. The bulk 
of the stars seen on the CMD is in the RC. 

Whenever we model the SFH of a galaxy, we first ask whether there are any features
that force us to assume a variable SFR over time. Therefore, in Fig.~12, we first show
a model with a constant SFR from 12~Gyr ago until the present. We then present two types 
of models with variable SFRs with time which are rooted in two distinct hypotheses. 
The basic assumption of Model~1 (Fig.~13a) is that there are no BHB stars in Leo~A. 
This is the model which will be most readily comparable with the result of Tolstoy et al. 
(1998). Model~2 (Fig~13b) investigates how much of an ancient BHB can be hidden in
the data, and was designed to contain the maximum number of BHB stars allowable by the data.

We start by describing the fiducial model with 12~Gyr, constant SFR shown in Fig.~12. 
For reference, such a model has a SFR of about 3.17 to 3.36 x 10$^{-5}$ M$_\odot$yr$^{-1}$. 
In the region  I$<$24, -1.0$<$(V-I)$<$0.5, this model produces more MS stars 
than the data, and has a distinct BL which the data lack as well. There are very few stars 
involved in producing the upper MS and these BL stars, hence we are limited by small-number statistics. 
On the other hand, it is easy enough to dis-allow very recent SF (so that the top of the MS becomes fainter),
and to reduce the SFR which gives rise to the bluest BL stars. 
In the region I$<$24,(V-I)$>$0.5, the model does quite well in producing the
correct number of stars in the upper RGB, and it is also successful in 
providing the observed number of stars in the RC as well as the mean RC brightness. There is
a spur of BL stars near (V-I)$\approx$0.6, which connects to the top of the RC.
Dis-allowing SF with ages near 1~Gyr avoids this ``blue finger" morphology. 
In the region  26$>$I$>$24, -1.0$<$(V-I)$<$0.5 the
model produces significantly too many stars in the blue plume. There is no indication of
the blue stars between the RC and the MS, which we could interpret as BHB stars.
However, the number of stars observed here is small, so this difference could be attributed to statistics. 
In the region (V-I)$>$0.5, there are significantly too few red, RGB stars below the RC. 
We therefore find that we cannot achieve a good match of the data with
a model that assumes constant star formation over the past 12~Gyr, and must look to a
time-variable SFR.

Next, we take inspiration from Tolstoy et al., who found that they needed to produce
at least 60\% of the stars on their CMD in a short burst of star formation with ages
from 0.9 to 1.5~Gyr. In Fig.~12, we illustrate the effect of such a SFH in our data.
It generates a completely different RC morphology from the one we observe. The RC is round,
not tilted; its mean brightness is too high. Furthermore, the upper RGB is blue and quite narrow.
(The same morphology appears when we assume the Tolstoy et al. distance modulus, but the
RC becomes even brighter and the upper RGB, even bluer). There are very few stars on the
RGB below the RC, where our data show a significant component.
Therefore, a better model SFH requires ages which reproduce correctly the mean location of the
RC, and also populates the lower part of the RGB/AGB without overpopulating its
upper part above the RC. The model discussed here also serves to
indicate that the spur of stars which we believe to be the signpost of a potential BHB component 
cannot be interpreted as the SGB of a burst of star formation that occured about 1.1~Gyr ago. 
This would bring along too many stars in the top blue part of the RC.
We conclude that we demonstrated with the help of synthetic CMDs our earlier inference that 
the SFH in the outer regions has clearly been different from that in the inner regions of Leo~A.

A solution for our ``no BHB" hypothesis is shown in Fig.~13a, top right. We first describe qualitatively,
the ingredients and constraints which went into this model. We did not produce any
very young stars in order to avoid making stars which are too luminous compared with the data, but we provide an
upper limit below. Any SFH which would produce most of the stars in
the blue plume with a young stellar component brought along too many BL stars. Therefore,
we introduced a gap in the SFH at around 1~Gyr and looked to the intermediate-age population to provide
additional stars for the bottom of the blue plume.
Any model that produced the required number of RC stars also brought along RGB/AGB stars, as well as
SGB stars which populate the bottom of the blue plume. Therefore, there is some crosstalk between
the intermediate-age and the recent SF. A synthetic RC aged 2---3 Gyr populated
the bottom of the blue plume very well, but the mean luminosity of the RC was still too
bright compared with the data (see Fig.~12). We also could not produce an RC with a simple, continuous SFR 
from 2---10~Gyr, and at the same time, provide a sufficient amount of stars to fill in the 
bottom of the CMD (see Fig.~12). 

Our best ``no BHB" model is qualitatively rather similar 
to that of Tolstoy et al. (1998).
We first tried to produce as young an RC as the data would allow. This yielded an ample amount
of SGB stars in the bottom of the blue plume, and we re-iterated on the SFH that produced
the upper part of the blue plume accordingly.
In this fashion we successfully synthesized an appropriate number of BL stars, and an acceptable luminosity
function for the blue plume. The young RC solution, however, fell short of synthetic
stars in the the bottom half of the RC, of RHB stars, and of stars on the RGB/AGB. These stars were
provided in the next step. Adding this third component allowed us to accomplish
several things. We found that it was possible to make these stars old enough
to avoid significant crosstalk with the blue plume, while bringing along enough stars in the bottom part
of the RC, and in the RHB. Also, this component provided stars on the RGB/AGB, which, in the
region above the RC, have red colors. This improved how well the model reproduces the overall
appearance of the RGB/AGB. 

Fig.~13a, bottom, illustrates what kind of Model~1 SFHs are compatible with the data. There
is no star formation in the recent 0.15~Gyr. As stated before, we suffer from small
number statistics here. We could easily allow a SFR of 5 x 10$^{-6}$ M$_\odot$yr$^{-1}$ in recent times.
The model has a young component with ages between 0.15 and 0.8~Gyr which has a SFR between 3.13 and
3.87 x 10$^{-5}$ M$_\odot$yr$^{-1}$, an intermediate-age component with an age range
from 1.7 to 4.5~Gyr with a SFR of 6.22 to 7.67 x 10$^{-5}$ M$_\odot$yr$^{-1}$, and an
old component with ages from 10 to 14~Gyr for which the SFR is between 
5.98 to 6.41 x 10$^{-5}$ M$_\odot$yr$^{-1}$. 
Although the SFR during the age interval 0.8 -- 1.7 Gyr is zero for the
example in Fig.~13a, we cannot rule out star formation at the level of a
few times 10$^{-6}$ M$_\odot$yr$^{-1}$ during this interval.
By pushing part of the RC to young ages, and part of
the RC to old ages, we produced a third gap in the SFH. It is quite likely that the
gap is not real. Instead of two epochs of SF separated by a zero SFR from 4.5 to 10~Gyr, slow modulations
in the intermediate and old SFR could produce similar CMDs. The possible parameter space for
solutions is large, and we did not explore it exhaustively. Stars with a wide range of masses and hence ages 
overlap in the area of the CMD which encompasses the RC; this introduces considerable degeneracy. 

We again find that the models suggest stars in  more outlying regions of Leo~A have higher mean ages. 
The young RC in the outer regions has an age of about 3~Gyr, compared to 1.2~Gyr in the inner regions
(Tolstoy et al.). In Model~1, about 30\% of the synthetic stars on the CMD are in the ancient component. 
In terms of the astrated mass, the ancient stars of Model~1 account for 1.14 times that in 
young and intermediate-age stars. 

Fig.~13b, top right, shows a solution for the ``maximum BHB" hypothesis. There are considerable
degeneracies in the model. Any event that produces
BHB stars populates the middle and bottom part of the blue plume, and thus
interferes with the young stellar component here, as well as with 
intermediate-age SF that populates SGBs near the bottom of the blue plume. 
We therefore started this model with the box that contains the BHB stars. Being
our ``maximum BHB" model, we tried to account for as many blue plume stars as possible 
with this component. This event also brought along a population of RHB
stars and RGB/AGB stars. By adjusting the duration of the SF event, we could
find a pleasing morphology for the HB. It was much simpler than in Model~1, to then also
find a good solution for the RC, since the RHB stars at the bottom of the RC, and some RGB/AGB 
stars were already accounted for. It was also easier with Model~2, to fill in the rest of the blue plume 
without overproducing BL stars. As intended, Model~2 yields copious stars
with I $\approx$ 25 between the two plumes where Model~1 shows none. 

Model~2 has a young component with ages from 0.05 to 
0.55~Gyr and SFRs between 2.85 and 3.07 x 10$^{-5}$ M$_\odot$yr$^{-1}$, an 
intermediate-age component aged 1.5 to 7.5~Gyr and
with SFRs in the range from 4.63 to 5.06 x 10$^{-5}$ M$_\odot$yr$^{-1}$, plus an ancient
stellar content with ages from 19.5 to 24~Gyr and SFRs between 5.92 to 6.67 x 10$^{-5}$ 
M$_\odot$yr$^{-1}$. In comparison with the inner regions of Leo~A as modeled by Tolstoy et al., Model~2
suggests a much higher mean age, of about 5.25~Gyr, for the bulk of the stars seen
on the CMD of the outer regions. The ancient population of Model~2 accounts 
for 20\% of the stars on the CMD, and its mass is 0.93 times that in young
and intermediate-age stars. 

As was anticipated from the isochrone comparison in the previous section,
the assumption of the presence of BHB stars requires very low-mass stars, which go along
with very high ages in our database. Specifically, the ages needed 
to produce a significant BHB are well in excess of the age of the Universe. 
There is an alternative interpretation, which would
make our Model~2 a more viable solution, and that is to allow for more 
mass loss on the RGB (see Lee, Demarque \& Zinn 1994). Our models
assumed Reimer's (1975) formula with $\eta$=0.3 (the canonical parameter
calibrated on the average properties of Galactic halo GCs). In this way, a star
loses about 0.1~M$_\odot$ on the RGB. If stars lose mass more efficiently 
on the RGB before they populate the HB, then low HB masses map onto younger ages.
For example, re-scaling the ages
of the stars which populate the BHB in our models for a mass loss of 0.2~M$_\odot$, shows 
their median age drops from 21.7~Gyr, to 11~Gyr! The apparent 
gap in SFR between 7.5 and 19.5~Gyr may therefore be considered an artifact of the RGB mass loss we choose; 
an increased mass loss on the RGB would allow for a younger absolute
age of the onset of SF, as well as for a different modulation from that of Model~2,
in the SFR at intermediate and old ages.

In summary, we find that the SFR in the outer regions of Leo~A has been very low 
in the most recent times, $<$0.1~Gyr ago, and was also very small 
around about 1~Gyr ago. We cannot produce a synthetic CMD which matches the data if
the bulk of the stars is between 0.9---1.5~Gyr old. 
We also cannot successfully model the data without allowing any
stars with ages well beyond 2~Gyr. As was clear from the isochrones, the presence of a RHB
demands stars with comparatively old ages. Just how old these stars are cannot be determined.
In Model~1, we produced a relatively young, 3~Gyr luminous RC, and combined it
with a 12~Gyr RC/RHB. In Model~2, we used a 5.25~Gyr RC and combined it with a
21.75~Gyr HB. Both models yield acceptable solutions in terms of CMD morphologies
and luminosity functions. Both models indicate an overall declining SFR from
early times to today.

The exploration of the SFH of Leo~A with synthetic CMDs based on the Z=0.0004 old Padua 
stellar evolution database has allowed us to gain additional insights into the
temporal and spatial dependence of the SFR in Leo~A. A more complete study
of the SFH of Leo~A, using both the dataset of Tolstoy et al. (1998) as well
as our data, and analyzed with the same techniques and assuming the same distance 
modulus, is desirable, but beyond the scope of the present work. In particular,
Castellani et al. (2000) present arguments that the new Padua tracks by Girardi et al. 
are preferred to simulate the RC. But Tolstoy et al. (1998) used the old Padua models,
and so did we for the simulations presented in this paper.
We also did not explore any metallicity evolution with the present simulations. 
If ongoing SF enriched the ISM of Leo~A, then metallicity changes with time are a likely candidate 
for widening the V-I color of the RGB/AGB. Here, its width was produced
with age differences, only. Conceivably, the stars on the BHB could belong to a different SF episode
than the RHB stars; if metallicity evolution is important the BHB stars could also have 
different metallicites from those on the RHB. The model-SFH of Leo~A and its total
astrated mass depend sensitively on how much mass loss is assumed on the RGB.

The simulations verify the results of the isochrone comparison by indicating
that Leo~A is an unlikely candiate for a delayed-forming
dwarf galaxy. The intriguing possibility of the detection of a BHB
should encourage deeper observations at larger galactocentric radii covering a wider field of view. The
measurement of ancient MSTOs, in particular, is needed to unambiguously confirm
the existence of a history of early star formation.

\section {Conclusions \& Implications}

We present deep B, V, \& I single-star photometry of an off-center
WFPC2 field in Leo~A, and use CMDs to discover the presence of metal-poor RHB stars. 
From this detection alone we can conclude that Leo~A contains some stars which 
have ages similar to the equivalent stars in Galactic GCs. Finding stars with old 
ages validates the use of the TRGB as a distance indicator to Leo~A. There is a 
significant component of young stars in our field, especially on the WF2 chip, 
and to a lesser degree, on the WF3 \& WF4 chips as well.
We cannot discriminate BHB stars from this young stellar
content, but present synthetic CMDs of WF3 \& WF4, which indicate the 
presence of BHB stars is possible. 

How old are the RHB stars which we see?
The absolute ages of GCs depend on their distances, but
the relative ages of metal-poor GCs which contain a RHB can be interpreted to
indicate that they could be about 1---2~Gyr younger than the oldest GCs (e.g. Lee et al. 2001). 
Assuming that the oldest clusters are between 11 and 13~Gyr old (Reid 1999), then a
reasonable lowest age limit for GCs with a metal-poor RHB is 9~Gyr. 
Therefore, Leo~A, showing a RHB and a metal-poor population, is at least 9~Gyr old.

We find that with a distance modulus of 24.5$\pm$0.2 on a short
RR Lyrae distance scale,
we can simultaneously account for the I-band magnitude of the TRGB
in ground-based data, and for the mean I-band magnitude of the RC in 
our WFPC2 field. A comparison with GC 
ridgelines indicates that the RGB has [Fe/H] smaller than 1\% of
Solar, which is below the O abundance of the ionized ISM, 3\% of Solar. 
Theoretical isochrones from a wide range of databases can account for
the locations in color and magnitude of the RGB/AGB branches, and
of the RC/RHB of Leo~A, when the stellar metallicities are assumed 
to be between 0.0001$<$Z$<$0.0004 (0.5-2\% of Solar)
and the stellar ages are between 5 and 10~Gyr. 

We use synthetic CMDs to evaluate the mean age of the RC stars. In the CMD of the
inner regions as modeled by Tolstoy et al. (1998), the RC has a mean age of about 1.2~Gyr.
The bulk of the stars which populate the RC on the CMD of the outer regions is 3~Gyr old
in our Model~1, and 5.25~Gyr old in our Model~2. Additionally, about half of the astrated
mass of either set of models resides in an ancient stellar population. The exact age of this
population can be adjusted depending on just how much mass loss is allowed on the RGB. 
The synthetic CMDs thus suggest a stellar substratum/halo population older than 9~Gyr 
is not inconsistent with the data, but its confirmation requires future deep, 
wide-field imaging further out in the halo of Leo~A.

Our findings have a variety of implications. Leo~A joins the ranks of other star-forming
dwarf galaxies which show a population gradient. While the mean age of the central 
regions of Leo~A is predominantly young, the mean age progressively increases with 
increasing galactocentric distance. The youngest MS, BSG, and BL stars 
are more concentrated toward the center;
their numbers fall off more rapidly with distance than those of the 
RGB stars which are present throughout the entire galaxy. 

Leo~A is not a young galaxy in the sense that it has not made
its very first stars in the past 2~Gyr. Our data
clearly show the signatures of old, metal-poor
stars. The cosmic SFR density has a peak at a redshift of z$\approx$1.5 (e.g. Boselli et
al. 2001, Fig.~8, Hopkins et al. 2001, Fig.~1).
Depending on the cosmological model, this redshift corresponds to lookback times of about
6.5---8.5~Gyr (see Hopkins et al., Figs. 3 and 4). Our limiting age of $>$9~Gyr for the oldest 
stars detected, indicates that the first stars in Leo~A were in place before the
cosmic SFR density started to decline. Our findings therefore add to the increasing evidence that some 
stars formed early on, 9~Gyr ago or earlier, in a wide range of Local Group galaxies. 

We see in Leo~A an extended history of star formation spanning billions of 
years. As with many other local dwarfs with deep CMDs, Leo~A
cannot readily be identified as a delayed-formation dwarf. 
The delayed-formation-of-dwarfs-scenario blossomed when it seemed to be
able to explain the excess of faint blue galaxies (e.g., Babul \& Ferguson 1996). 
We have since learned that the highest
SFRs observed in the fossil record of local dwarfs (which typically allow
a detailed construction of their star formation history over the past $\sim$~Gyr)
are rarely ever high enough to account for the blue luminosities required of
such briefly bursting dwarfs, and that the time-scales for star formation
are usually longer than the 10~Myr assumed in the Babul \& Ferguson scenario
(Greggio et al. 1998, Lynds et al. 1998, Schulte-Ladbeck et al. 2001, Tosi 2001). Therefore, such
an extreme bursting mode seems unlikely.  Furthermore, 
the direct observation of (potentially) high 
metallicities, i.e., oxygen abundances, in faint blue galaxies
suggests that they are typically higher mass systems than local dwarfs (Carollo \& Lilly 2001).

The discovery of a young galaxy in the local Universe would have profound
implications for theories of galaxy formation, and this possibility has
long motivated searches for galaxies with very low oxygen abundances and
very blue colors (Sargent \& Searle 1970; Izotov \& Thuan 1999;
Kunth \& \"{O}stlin 2000). The existence of old stars in Leo~A, however,
joins mounting evidence that star-forming dwarf galaxies in the
local Universe do contain old stars, no matter how low the
metallicity of their ionized gas.  In other words, an extremely 
low oxygen abundance does not imply a recently formed galaxy.

\acknowledgments

We thank Dr. E.~Tolstoy for supplying us with her photometry
of the inner regions of Leo~A. We are grateful 
to Claus G\"{o}ssel for obtaining the R-band image of Leo~A at
the Wendelstein Observatory of the University of Munich.
We would also like to thank Dr. C.~Maraston for sharing with
us her set of Frascati isochrones. We made extensive use of
the SIMBAD and NED databases. This work was supported by
an HST grant associated with GO program 8575.  
UH would also like to acknowledge financial support from SFB~375.

\clearpage

\vspace{-1.5cm}

\figcaption[sgi9259.eps]{The locations (WFPC2 footprints) of the Tolstoy et al. data (GO program
5915) and of our data (GO program 8575) relative to a Leo~A R-band image 
obtained at the Wendelstein Observatory. The WFPC2 insets show the resolved stars
in the V filter, and have been scaled relative to one another based on total
exposure time. Two WF chips side-by-side have a length of 154''; this provides a sense of
scale for this image. \label{fig1}}

\figcaption[sgi9259.eps]{a. The CMD of resolved stars in each of the WFPC2 chips
for the first filter combination. \label{fig2 a}}

\noindent Fig. 2b. -- The CMD of resolved stars in each of the WFPC2 chips
for the second filter combination.

\noindent Fig. 2c. -- The CMD of resolved stars in each of the WFPC2 chips
for the third filter combination.

\figcaption[sgi9259.eps]{DAOPHOT photometric errors for all three filters 
on the WF3 and WF4 chips (magnitudes transformed to B, V, I). \label{fig3}}

\figcaption[sgi9259.eps]{Completeness tests from ADDSTAR for all three filters 
on the WF3 and WF4 chips (magnitudes transformed to B, V, I). \label{fig4}}

\figcaption[sgi9259.eps]{a. Luminosity functions along the blue plume in F555W$_0$.
The error bars show the squareroot of the starcounts in each bin.
Notice the dominance of the MS on WF2. In the more outlying fields, WF3 and WF4,
the MS is much weaker.\label{fig5 a}}

\noindent Fig. 5b. -- Luminosity functions along the red plume in F814W$_0$.
The error bars show the squareroot of the starcounts in each bin. 
Notice how the red clump, located at F814W$_0$ $\approx$ 24, has a smaller
FWHM in the WF~3 and WF4 data compared with the PC1 and WF2 data.

\figcaption[sgi9259.eps]{Top: The V, I CMD of all stars on all chips. Bottom:
I-band luminosity function along the top of the red plume. The error bars shown reflect 
the squareroot of the starcounts in each bin. The
magnitude of the TRGB from Tolstoy et al. (1998) based on ground-based, wide-field imaging
is shown by the dashed line. \label{fig6}}

\figcaption[sgi9259.eps]{The V, I CMD of all stars with errors smaller than 0.2~mag.
The Globular Cluster ridgelines of Da Costa \& Armandroff (1990) are overlayed.
I$_{TRGB,0}$ = 20.5 (see the dashed line) and (m-M)$_0$ = 24.5. 
The GCs and their metallicities, [Fe/H],
are M~15, -2.17, NGC~6397, -1.91, M~2, -1.58, NGC 6752, -1.54, NGC~1851, -1.29, and
47~Tuc, -0.71.  \label{fig7}}

\figcaption{The I-band luminosity function in the neighborhood of the red clump.
The error bars shown reflect the squareroot of the starcounts in each bin. 
A fit which combines a linear part for the RGB and a Gaussian for the RC is
overplotted. From this we determine I$_{RC,0}$ = 24.05. \label{fig8}}

\figcaption[sgi9259.eps]{Left: [(V-I)$_0$, I$_0$] CMDs of all data with
errors smaller than 0.1~mag in V and I. Right: [(B-I)$_0$, I$_0$] CMDs of all data with
errors smaller than 0.1~mag in B and I. The dotted lines at
I$_0$ = 20.5 mark the observed location of the TRGB. The isochrones were matched to
the observed TRGB (but since the AGB phase extends above the
TRGB, this is not obvious from the plots). The data are overplotted with Z=0.0004
old and new Padua isochrones. The top panels show isochrones of young ages,
the middle panels an isochrone of an intermediate age, and the bottom panels
compare the data with a very old isochrone.  \label{fig9}}

\figcaption[sgi9259.eps]{The [(B-I)$_0$, I$_0$] CMD of Leo~A. Only stars with
errors of up to 0.1~mag are plotted. The dotted lines at
I$_0$ = 20.5 mark the observed location of the TRGB. Yale isochrones for
three ages and two metallicities are overplotted onto the data. They illustrate
the age-metallicity degeneracy of the RGB. If the metallicity of the RGB stars
is as high as that of the ISM of Leo~A, then the age of the bulk of RGB stars is around
$\approx$5~Gyr. Model RGB stars with metallicities slightly below that of stars in M~15
(the GC which empirically describes the Leo~A RGB) 
are slightly too blue, even for the oldest ages. \label{fig10}}

\figcaption[sgi9259.eps]{The [(B-I)$_0$, I$_0$] CMD of Leo~A. Only stars with
errors of up to 0.1~mag are plotted. The dotted lines at
I$_0$ = 20.5 mark the observed location of the TRGB. Padua isochrones for
three ages and two metallicities are overplotted onto the data. Frascati
isochrones for Z=0.0001 are also overplotted for three ages. The RC/RHB requires
stars no older than about 10~Gyr. The Frascati isochrones
can match the RGB/AGB as well as the RC/RHB. \label{fig11}}

\figcaption[sgi9259.eps]{Top: The [(V-I$_0$, I$_0$] CMD of stars on the WF3 \& WF4 chips.
A few dashed lines are drawn to help compare the data with two models below.
Middle: The baseline synthetic V, I CMD which results if we simply assume a constant SFR
starting 12~Gyr ago and continuing on to the present. The differences between this
model and the data guide us to a SFH which varies with time. Bottom:  The SFH which
produced most of the stars on the V, I CMD of the inner regions (Tolstoy et al. 1998)
generates a red clump and giant branch morphology which is inconsistent with
our data of the outer regions.\label{fig 12}}

\figcaption[sgi9259.eps] {a.Top left: The [(V-I$_0$, I$_0$] CMD of stars on the WF3 \& WF4 chips.
Top right: The ``best" synthetic V, I CMD that results from Model~1 (``no BHB"). Middle: Comparison of the 
data and the model in terms of their I-band LFs. Bottom: The SFH used to synthesize Model~1.
The ``zero" SFRs before the first, and between the first and the second event
are real, in the sense that we do believe that the SFRs here were much smaller than
either in the first or in the second event. The gap in SFR between the second and
the third event is ill constrained.\label{fig 13a}}

\noindent Fig. 13b. --  Top left: The [(V-I$_0$, I$_0$] CMD of stars on the WF3 \& WF4 chips.
Top right: The ``best" synthetic V, I CMD that results from Model~2 (``maximum BHB"). Middle: Comparison of the 
data and the model in terms of their I-band LFs. Bottom: The SFH used to synthesize Model~2. Again,
the gap in SFR between the second and the third event is ill constrained.

\clearpage
\section {Appendix A --- Comparison with M~15}

The purpose of this Appendix is twofold. First, a large body of data on GCs exists 
in B and V. We did not tap into this information in the main body
of the paper, because we used the I-band TRGB to pin down the distance. 
Second, we wish to make a few remarks on the cluster distance scale, 
and how it affects the interpretation of Leo~A data.   

In Fig.~A1, we show our best match of the M~15 [(B-V)$_0$, V$_0$] CMD to the Leo~A [(B-V)$_0$, V$_0$]
CMD (in an attempt to imitate the main-sequence-fitting technique). 
We used only the outlying fields of Leo~A for this plot because the young component which overlaps
a potential BHB is much weaker here. In Fig.~A1, the solid lines for M~15 represent 
the data of Sandage (1970). The lines
show the locations of the RGB, AGB, and HB of M~15 when shifted onto the Leo~A data. 
The shift applied is 9.47~mag. This yields an excellent match to the RGB, AGB, RHB
of Leo~A. It also shows that some of the stars which are to the blue
of the RGB can be interpreted as metal-poor AGB stars. 
Because the Sandage data are based on rather old, photoelectric photometry, we
also constructed ridgelines using the CCD photometry of Durrell \& Harris (1993). These
are overplotted as dashed lines (after the same shift was applied). We see that
there is about a 0.25~mag offset at the TRGB between the two data sets, but the overall
fit is good. We also indicate where the RR~Lyrae stars of M~15 (Silbermann \& Smith 1995) are located 
when they are shifted by 9.47~mag\footnotemark[4]. These comparisons indicate that the distance 
modulus of Leo~A is about 9.5~mag larger than that of M~15. 

\footnotetext[4]{Silbermann \& Smith give a mean apparent V magnitude for the
RR~Lyr stars in M~15 of 15.82$\pm$0.03. Assuming an A$_V$ of 0.31, then their dereddend V magnitude is 15.51.
Dolphin et al. (2002) now find RR~Lyr variables in Leo~A with a mean V magnitude of 25.10$\pm$0.09. 
We assumed this was not extinction corrected, and applied an A$_V$ of 0.068. We see that 
the difference in the magnitudes of the RR~Lyr variables in M~15 and in
Leo~A is 9.52, in excellent agreement with the value we derive from shifting M~15's RGB, AGB, and HB
onto the Leo~A CMD.}

Hipparcos observations of local subdwarfs have been used to re-derive the distances to
Galactic GCs (Reid 1997, Reid \& Gizis 1998, Reid 1999). We notice that one effect of the
new GC distances is to shift the absolute I-band TRGB magnitudes of Da Costa \& Armandroff (1990), 
which are commonly used to derive distances to external galaxies. These shifts 
are not in concert. Relative
shifts in GC distances conspire to introduce a larger dispersion of M$_I$ at the TRGB than 
the pre-Hipparcos distances had indicated. With the post-Hipparcos distances to GCs, 
the error associated with the TRGB method increases. (We estimate that the dispersion is now 0.2, 
rather than 0.1~mag, in M$_I$. But note that Reid (1999) gives an uncertainty 
of at least $\pm$0.1~mag in post-Hipparcos cluster distance moduli.) Similarly, since
Da Costa \& Armandroff calibrate metallicity against V-I color, metallicities and metallicity
errors derived based on GC ridgelines change when the new cluster distances are adopted.
More importantly, however, the absolute distance modulus of M~15 has been re-calibrated in
the post-Hipparcos era. Reid (1997) gives (m-M)$_0$=15.38, while Da Costa \& Armandroff (1990)
assumed (m-M)$_0$=15.10. Therefore, the Leo~A distance modulus is 24.9 using the new M~15 distance,
and 24.6 using the old M~15 distance.

In principle there are two sources of error for the distance modulus coming from the calibration,
once M~15 has been chosen as the standard candle.
One is related to the relative modulus between M~15 and Leo~A; the other
comes from the modulus adopted for M~15. Reid (1999) quotes an uncertainty 
of at least $\pm$0.1~mag in the cluster distance moduli. The error in the
relative modulus between M~15 and Leo~A has two components. One arises from the  
difference between the two M~15 data sets at the TRGB, $\pm$0.13~mag, which most
probably results from different sampling at the TRGB. Uncertainty in the sampling
of the Leo~A TRGB adds at least another 0.1--0.2~mag.
Therefore, the total error on the distance modulus of Leo~A as derived from
comparison with M~15 is $\approx$$\pm$0.2.

The large distance to Leo~A which results when one adopts the new distance of M~15
does not change our main conclusions. We detect old 
and metal-poor RHB stars.  Therefore, Leo~A cannot be younger than 2~Gyr. 
However, the material presented in
this Appendix serves to illustrate the difficulty in disentangling distance, metallicity, 
and age from one another, when neither one is known with certainty. 

\clearpage

\noindent Fig. A1. -- The [(B-V)$_0$, V$_0$] CMD of the WF3 \& 4 chips. Only stars with
errors in B and V of up to 0.1~mag are plotted. We overplot two ridgeline sets for
M~15. The solid lines give the RGB, AGB and HB after 
Sandage (1970), shifted by 9.47~mag. To verify the locations with modern data we also constructed 
an RGB and a BHB ridgeline from the data of Durrell \& Harris (1993); these are 
the dashed lines. The small cross shows the location of RR~Lyrae variables in M~15 
after Silbermann \& Smith (1995).


\begin{references}

\reference{ } Aloisi, A., Tosi, M., Greggio, L. 1999, AJ, 118, 302
\reference{ } Babul, A., Ferguson, H.C., 1996, ApJ, 458, 100
\reference{ } Barkana, R., Loeb, A. 2001, astro-ph/0010468, to be published
in Physics Reports 2001
\reference{ } Ballazzini, M., Ferraro, F.R., Pancino, E. 2001, ApJ (astro-ph/0104114)
\reference{ } Bessel, M.S., Castelli, F., Plez, B. 1998, A\&A, 1998, 337
\reference{ } Bertelli, G., Bressan, A., Chiosi, C., Fagotto, F., Nasi, F. 
1994 A\&AS, 106, 275
\reference{ } Boselli, A., Gavazzi, G., Donas, J., Scodeggio, M. 2001, AJ, 121, 753
\reference{ } Caputo, F., Castellani, V., Del'Innocenti, S. 1995, A\&A, 304, 365
\reference{ } Carollo, C.M., Lilly, S.J. 2001, ApJ, 548, L153
\reference{ } Carretta, E. Gratton, R.G., Clementini, G., Fusi Pecci, F. 2000, ApJ, 533, 215
\reference{ } Castellani V., Degl'Innocenti, S., Girardi, L.,
 Marconi, M., Prada Moroni, P. G., Weiss, A. 2000, A\&A, 354, 150
\reference{ } Castelli F., Gratton, R.G, Kurucz, R.L. 1997a, A\&A, 318, 841
\reference{ } Castelli F., Gratton, R.G, Kurucz, R.L. 1997b, A\&A, 324, 432
\reference{ } Castelli F., Degl'Innocenti, S., Girardi, L., Marconi, M., Prada Moroni,P.G., Weiss, A.
2000, A\&A, 354, 150 
\reference{ } Cassisi S., Castellani, M., Castellani V. 1997, A\&A 317,108
reference{ }  Cole, A.A., Tolstoy, E., Gallagher, J.S. et al. 1999, AJ, 118, 1657
\reference{ } Cotton, W.D., Condon, J.J., Arbizzani, E. 1999,ApJS, 125, 409
\reference{ } Crone, M.M., Schulte-Ladbeck, R.E., Greggio, L., Hopp, U. 2002, ApJ, 567, 258
\reference{ } Da Costa, G.S., Armandroff, T.E. 1990, AJ, 100, 162
\reference{ } Demers, S., Kibblewhite, E.J., Irwin, M.J., Bunclark, P.S., Bridgeland, M.T.
1994, AJ, 89, 1160
\reference{ } Dekel, A., Silk, J. 1986, ApJ, 303, 39
\reference{ } Dolphin, A.E., Saha, A., Skillman, E.D., Tolstoy, E.,
Cole, A.A., Dohm-Palmer, R.C., Gallagher, J.S., Mateo, M., Hoessel, J.G. 2001, ApJ,
550, 554
\reference{ } Dolphin, A.E., Saha, A., Claver, J., Skillman, E.D., Cole, A.A., Gallagher, J.S.,
Tolstoy, E., Dohm-Palmer, R.C., Mateo, M., 2002, AJ, in press, astro-ph/0202381
\reference{ } Duc, P.-A., Mirabel, I.F. 1998, A\&A, 333, 813
\reference{ } Durrell, P.R., Harris, W.E. 1993, AJ, 105, 1420
\reference{ } Ellis, R.S. 1997, ARA\&A, 35, 389
\reference{ } Fagotto, R., Bressan, A., Bertelli, G., Chiosi, C. 1994, A\&AS, 105, 29
\reference{ } Ferrara, A., Tolstoy, E. 2000, MNRAS, 313, 291
\reference{ } Girardi, L., Bressan, A., Bertelli, G., Chiosi, C. 2000 A\&AS, 141, 371
\reference{ } Girardi, L., Salaris, M. 2001, MNRAS, 323, 109
\reference{ } Greggio, L., Tosi, M., Clampin, M., De Marchi, G.,
Leitherer, C., Nota, A., Sirianni, M. 1998, ApJ, 504, 725
\reference{ } Grebel, E.K., 1999, in IAU Sump. 192, The stellar Contnet of the Local Group,
eds. p. Whitelock \& R. Cannon (Provo: ASP), 17
\reference{ } Grebel, E.K., 2000, astro-ph/0008249, to appear in ``Microlensing 2000: 
A New Era of Microlensing Astrophysics", Cape Town, South Africa, ASP Conf.
Ser., ed. J.W. Menzies and P.D. Sackett
\reference{ } Harbeck, S., Brebel, E.K., Holtzman, J., Guuhathakurta, P., Brandner, W.,
Geisler, D., Sarajedini, A., Dolphin, A., Hurley-Keller, D., Mateo, M. 2001, astro-ph/0109121
\reference{ } Hesser, J. E., Harris, W. E., Vandenberg, D. A. 1987, PASP, 99, 1148
\reference{ } Hodge, P.W., Dolphin, A.E., Smith, T.R., Mateo, M. 1999, ApJ, 521, 577
\reference{ } Holtzman J.A., Burrows C.J., Casertano S., Hester J.J., Trauger J.T., 
Watson A.M., Worthey G., 1995, PASP, 107, 1065
\reference{ } Hopkins, A.M., Irwin, M.J., Connolly, A.J. 2001, ApJ, 558, L31
\reference{ } Izotov, Y.I., Thuan, T.X. 1999, ApJ, 511, 639
\reference{ } Izotov, Y.I., Chaffee, F. H., Foltz, C.B., Green, R.F.,
 Guseva, N.G., Thuan, T.X. 1999, ApJ, 527, 757
\reference{ } Klypin, A., Kravtsov, A.V., Valenzuela, O., Prada, F. 1999, ApJ, 522, 82 
\reference{ } Kunth, D., \"{O}stlin, G. 2000, A\&AR, 10, 1
\reference{} Lee, M.G., Freedman, W.L., Madore, B.F. 1993, ApJ, 417, 553
\reference{} Lee, Y.-W., Demarque, P., Zinn, R. 1990, ApJ, 350, 155
\reference{} Lee, Y.-W., Demarque, P., Zinn, R. 1994, ApJ, 423, 248
\reference{} Lee, Y.-W., Yoon, S.-J., Rey, S.-C. 2001, to appear in ``Astrophysical
Ages and Time Scales, ASP Conf. ser., edts. T. von Hippel, N. Manset, \& C. Simpson, 
astro-ph/0104405
\reference{} Lejeune, Th., Schaerer, D. 2001, A\&A, 366, 538
\reference{} Lynds, R., Tolstoy, E., O'Neill, E.J., Jr., Hunter, D.A. 1998, AJ, 116, 146
\reference{ } Marzke, R.O., Da Costa, L.N. 1997, AJ, 113, 185
\reference{ } Mateo, M. 1998, ARA\&A, 36, 435
\reference{ } Mateo, M. 2000, in ``The First Stars", Proceedings of 
the MPA/ESO Workshop held at Garching, Germany, 4-6 August 1999, Achim Weiss, Tom G. Abel, Vanessa
Hill (eds.), Springer, p.~283
\reference{ } Mighell, K.J., Sarajedini, A., French, R.S. 1998, AJ, 116, 2395
\reference{ } Minniti, D., Zijlstra, A.A. 1996, ApJL, 467, 13
\reference{ }\"{O}stlin, G. 2000, ApJL, 535, 99
\reference{ } Origlia, L., Leitherer, C. 2000, AJ, 119, 2018
\reference{ } Popowski, P., Gould, A. 1998, in ``Post-Hipparcos Cosmic Candels", A. Heck, F. Caputo
(eds.), Kluwer Academic Publ., Dordrecht, p. 53
\reference{ } Reid, I.N. 1997, AJ, 114, 161
\reference{ } Reid, I.N. 1999, ARA\&A, 37, 191
\reference{ } Reimers, D. 1975, in ``Problems in Stellar Atmospheres and Envelopes", ed. 
B. Baschek, W. H. Kegel, \& G. Traving (Berlin: Springer), 229
\reference{ } Rejkuba, M., Minniti, D., Gregg, M.D., Zijlstra, A.A., Alonso, M.V., Goudfrooij, P. 
2000, AJ, 120, 801
\reference{ } Rood R.T. 1973, ApJ, 184, 815
\reference{ } Roukema, B.F., Peterson, B.A., Quinn, P.J., Rocca-Volmerange, B. 1997, MNRAS, 292, 835
\reference{ } Roukema, B.F. 1998, in ``Dwarf Galaxies and Cosmology", Proceedings of the 33$^{rd}$
Moriond Astrophysics Meeting held at Les Arcs, France, Narch 14-21, 1998, ed. T.X. Thuan, C. Balkowski,
V. Cayrate, J. Tran Thanh Van, Editions Fronti\`{e}res, p. 471
\reference{ } Saha, A., Freedman, W.L., Hoessel, J.G., Mossman, A.E. 1992, AJ, 104, 1072
\reference{ } Salpeter, E.E. 1995, ApJ, 121, 161
\reference{ } Sandage, A. 1970, ApJ, 162, 842
\reference{ } Sargent, W.L.W., Searle, L.  1970, ApJL, 162, 155 
\reference{ } Schulte-Ladbeck, R.E., Hopp, U., Crone, M.M., Greggio, L. 1999, ApJ, 525, 709
\reference{ } Schulte-Ladbeck, R.E., Hopp, U., M.M., Greggio, L., Crone, M.M. 2000, AJ, 120, 1713
\reference{ } Schulte-Ladbeck, R.E., Hopp, U., Greggio, L., Crone, M.M., Drozdovsky, I.O. 2001,
AJ, 121, 3007
\reference{ } Silbermann, N.A., Smith, H.A. 1995, AJ, 110, 704
\reference{ } Skillman, E.D., Kennicutt, R.C., Hodge, P. 1989, ApJ, 347, 875
\reference{ } Stetson, P.B. 1994, PASP, 106, 250
\reference{ } Tolstoy, E., Gallagher, J.S., Cole, A.A., Hoessel, J.G., Saha, A., Dohm-Palmer, R.C., 
Skillman, E.D., Mateo, M., Hurledy-Keller, D. 1998, AJ, 116, 1244
\reference{ } Tosi, M. 2001, review paper to appear in 'Dwarf Galaxies and their Environment", 
K.S. de Boer, R.J. Dettmar, U. Klein eds. (Shaker Verlag, De), astro-ph/0104016
\reference{ } Tosi, M., Sabbi, E., Bellazzini, M., Aloisi, A., Greggio, L., Leitherer, C.,
\& Montegriffo, P. 2001, AJ 122, 1271
\reference{ } Udalski A. 2000a, Acta Astron., 50, 279
\reference{ } Udalski A. 2000b, ApJ, 531, L25 
\reference{ } Walker, A. 1989, PASP, 101, 570
\reference{ } White, S.D.M., Springel, V. 2000, in ``The First Stars", Proceedings of 
the MPA/ESO Workshop held at Garching, Germany, 4-6 August 1999, ed. A. Weiss, T. G. Abel, V.
Hill, Springer, p.~327
\reference{ } van Zee, L., Skillman, E.D., Haynes, M.P. 1999, AAS, 194.0504 
\reference{ } Yi, S., Demarque, P., Kim,  Y.-C., Lee, Y.-W., Ree, C., Lejeune, Th., Barnes, S.
2001, ApJ, in press (astro-ph/0104292)
\end{references}
\end{document}